Super-R BiFeO$_3$: Epitaxial stabilization of a low-symmetry phase with giant electromechanical response

Oliver Paull,[1,*] Changsong Xu,[2,*] Xuan Cheng,[3] Yangyang Zhang,[1] Bin Xu,[2,4] Kyle Kelley,[5] Liam Collins,[5] Alex de Marco,[3,6] Rama K. Vasudevan,[5] Laurent Bellaiche,[2] Valanoor Nagarajan,[1] and Daniel Sando[1,7]

[1]School of Materials Science and Engineering, UNSW Sydney, High Street, Kensington, 2052, Australia

[2]Department of Physics and Institute for Nanoscience and Engineering, University of Arkansas, Fayetteville, AR 72701, USA

[3]Department of Biochemistry and Molecular Biology, Monash University, Clayton, VIC 3800, Australia

[4]School of Physical Science and Technology, Soochow University, Suzhou 215006, China

[5]The Center for Nanophase Materials Sciences, Oak Ridge National Laboratory, Oak Ridge, TN 37831, USA

[6]ARC Centre of Excellence in Advanced Molecular Imaging, Monash University, Clayton, VIC 3800, Australia

[7]Mark Wainwright Analytical Centre, UNSW Sydney, High Street, Kensington, 2052, Australia

*these authors contributed equally

daniel.sando@unsw.edu.au

nagarajan@unsw.edu.au

Piezoelectrics interconvert mechanical energy and electric charge and are widely used in actuators and sensors. The best performing materials are ferroelectrics at a morphotropic phase boundary (MPB), where several phases can intimately coexist. Switching between these phases by electric field produces a large electromechanical response. In the ferroelectric BiFeO$_3$, strain can be used to create an MPB-like phase mixture and thus to generate large electric field dependent strains. However, this enhanced response occurs at localized, randomly positioned regions of the film, which potentially complicates nanodevice design. Here, we use epitaxial strain and orientation engineering in tandem – *anisotropic epitaxy* – to craft a hitherto unavailable low-symmetry phase of BiFeO$_3$ which acts as a structural bridge between the rhombohedral-like and tetragonal-like polymorphs. Interferometric displacement sensor measurements and first-principle calculations reveal that under external electric bias, this phase undergoes a transition to the tetragonal-like polymorph, generating a piezoelectric response enhanced by over 200%, and associated giant field-induced reversible strain. These results offer a new route to engineer giant electromechanical properties in thin films, with broader perspectives for other functional oxide systems.

Piezoelectrics interconvert mechanical energy and electric charge[1], and are exploited in a wide variety of applications such as switches, relays, actuators, and sensors[2]. In ferroelectrics (a subset of piezoelectrics which possess a switchable polarization), the electromechanical response peaks in the vicinity of a phase boundary between several crystallographic phases with similar free energy profiles[3]. At so-called morphotropic phase boundaries (MPBs) (refs. [4–6]), the symmetry of the crystal structure abruptly changes and is easily interconverted by external stimuli. Consequently, MPB-based ferroelectrics (such as (Pb,Zr)TiO$_3$ – PZT) demonstrate exceptional piezoelectric coefficients[6–12]. The exact origin of the enhanced piezoelectric response at MPBs remains a topic of debate; *e.g.* see refs. [11,13] *vs.* refs. [4,14]. Whether this linking phase is called "adaptive"[13] or "bridging low-symmetry"[15], undoubtedly at its core it must facilitate the straddling across the boundary.

In BiFeO$_3$ (BFO), a room-temperature multiferroic (materials with coexisting ferroelectric and magnetic orders) with rich ferroelectric[16], optical[17], and magnetic[18] phase diagrams, a "*strain*-driven MPB" was demonstrated[12]. Bulk BFO possesses rhombohedral *R*3*c* symmetry (R), but when synthesized as an epitaxial film under high compressive strain, it can crystallize in a tetragonal-like (T') phase with high axial ratio (c/a $\approx$ 1.23) (ref. [19]). This strain-driven transition is accompanied by a systematic polarization rotation from [111] (R) towards [001] (T) and a change in crystal symmetry (**Fig. 1a**). Increasing the BFO film thickness beyond ~25 nm triggers strain relaxation, yielding an MPB-like phase mixture comprising T' and strained rhombohedral-like (R') polymorphs[12]. In this "mixed-phase BFO", the R' and T' phases, which possess $M_A$ and $M_C$ (both monoclinic) symmetries respectively[16] (**Fig. 1a**), form characteristic striations interspersed randomly throughout the T' matrix.

Mixed-phase BFO possesses various attractive functionalities with implications for next-generation multifunctional materials systems[12]. Of particular relevance here is that an electric field applied locally to the mixed-phase regions induces a large, physical deformation of the sample[20,21], with concomitant enhanced electromechanical response, arising from a phase transformation between the R' and T' polymorphs[22].

In the monoclinic $M_A$ and $M_C$ phases, the polarization axis is constrained to a crystallographic *plane* (blue shaded regions in **Fig. 1a**), rather than a single crystallographic *direction* (for, *e.g.,* R and T phases, arrows in **Fig. 1a**)[15]. Therefore, at the R'/T' phase *boundary* in mixed-phase BFO, the polarization vector is contained in some plane that links the $M_A$ and $M_C$ symmetries. Since this plane corresponds to a high index surface, the phase necessarily has low symmetry (green shaded plane in **Fig. 1a**). To convert between the R' and T' phases, the most energetically favourable pathway is through a mechanically-soft[23] triclinic phase[24], reminiscent of the MPB model for PZT proposed by Noheda *et al.*[14].

A key feature of such a pathway is the concept of polarization rotation[25,26], which is enabled by the low crystallographic symmetry of the system, and widely associated with very large piezoelectric constants[1].

Indeed, ferroelectrics generally can be classified as either 'rotator' (where the polarization rotates in some plane) or 'extender' (the polarization lengthens along some high symmetry direction)[27]. In the BFO system specifically, in this context one would consider the 'rotator' description apt, as the various monoclinic phases (along with some intermediate triclinic phase) are characterised by the ability of the polarization vector to rotate in the relevant plane. In mixed-phase BFO, therefore, the enhanced electromechanical performance is arguably driven by polarization rotation that occurs in these mechanically soft low-symmetry phases within the mixed-phase striations[28].

In contrast with other MPB systems such as PZT or $Pb(Mg,Nb)O_3$-$PbTiO_3$ (PMN-PT) where the bridging phase can be obtained at the correct composition, in typical mixed phase BFO phase separation into the R' and T' constituents is unavoidable[16]. While the relative phase fractions can be optimized using thickness control[12], manipulating the free energy landscape of epitaxial BFO to gain access to the true bridging phase has remained elusive. If one could stabilize the soft triclinic polymorph as a monolithic *single* phase which is able to sustain homogeneous and reversible conversion to T', then a dramatically enhanced piezoelectric response could be expected.

Here, we use an approach called *anisotropic epitaxy* (AE) using the highly miscut (310) $LaAlO_3$ (LAO) substrate to demonstrate large-scale epitaxial stabilization of a highly strained, highly-distorted R' phase of BFO. Electromechanical measurements using an interferometric displacement sensor (IDS), combined with first-principles-based calculations, reveal that applied electric field induces homogeneous and reversible conversion from the distorted R' (*i.e.,* triclinic) phase to a T' (*i.e.,* $M_C$) phase. This phase transition is accompanied by a dramatic elastic softening and a giant increase in the electromechanical strain. Experimentally, we demonstrate a reversible strain of ~1% and abnormally large d'$_{33}$ coefficient (along the out-of-plane direction) approaching of 200 pm/V. Critically, as the enhanced electromechanical response does not originate from specific domain wall or phase boundary motion, but rather from homogenous interconversion of the entire sample area under the stimulus, it provides benefits for device applications. Moreover, since the interconversion is triggered at modest voltages (~5 V), this system is attractive for nanoscale electromechanical systems where the driving voltage is typically limited to values available with CMOS electronics.

**Establishing the basis for our approach**

In bulk ceramics and single crystals, MPBs are traditionally obtained through hydrostatic pressure[4], doping[29], intentionally-induced defects[30], and/or disorder[29]. In epitaxial thin films, however, the mechanical and electrical boundary conditions are highly tuneable, and can be used to engineer phase transitions[31]. Epitaxial strain engineering, for instance, conventionally applied under isotropic conditions, can stabilize novel ferroelectric domain arrangements and structural phases[32]. Various substrate crystallographic orientations, such as (100), (110), and (111), yield different film symmetries with prominent modifications in functional properties[33,34].

These traditional heteroepitaxy approaches are not capable of stabilizing, in monolithic form, the desired (and thus far elusive) distorted R' phase. This is since the distorted R' phase of BFO observed in mixed-phase samples has a strong crystallographic tilt and a lower (likely triclinic) symmetry[16]. Conventional strain engineering cannot yield such a distortion as a strong tilt-inducing anisotropy is not imposed by the substrate – thereby necessitating the use of anisotropic miscut substrates. Such miscuts have been used to grow low-symmetry triclinic BFO[35]; however, no information is available regarding its functional properties.

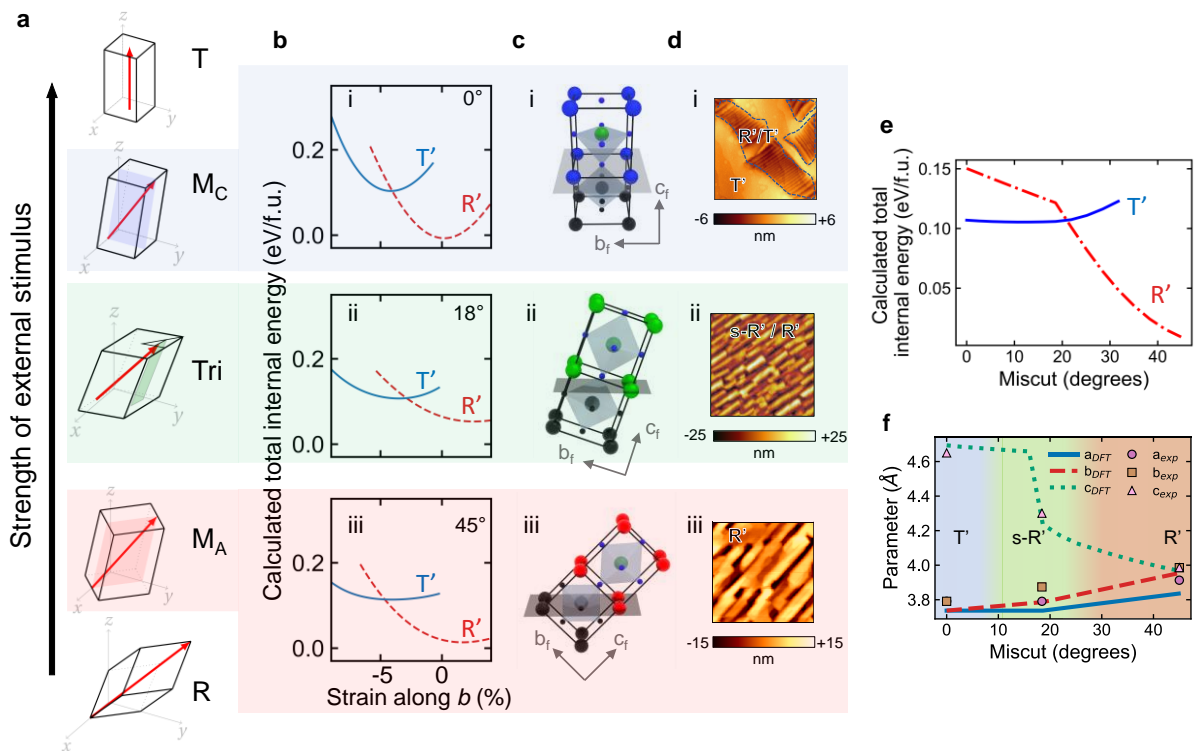

**Figure 1. Concept of anisotropic epitaxy to obtain exotic ferroelectric phases.** (**a**) Schematic of crystal symmetry in ferroelectrics with increasing external stimulus (composition, applied field, etc.). The red arrows denote the direction of the polarization. Note that for the lower symmetry monoclinic $M_C$ and $M_A$ phases, the polarization is confined to a plane (shaded blue and red respectively), while for triclinic, the polarization is confined to some plane (shaded green) between the $M_C$ and $M_A$ planes. (**b**) Calculated total internal energy *vs.* in-plane strain for the T' and R' phases at various miscut angles: (**i**) 0°, (**ii**) ~18.4°, (**iii**) 45°. (**c**) Perovskite pseudocubic unit cell for (**i**) T'-phase (0° miscut), (**ii**) super-R' phase (18.4° miscut), and (**iii**) R'-phase (45° miscut). (**d**) Corresponding topography (Asylum Cypher; contact mode) for $BiFeO_3$ films (**i**) (50 nm) on (001)-oriented (mixed-phase R'/T'), (**ii**) (20 nm) on (310)-oriented (mixed super-R' / R'), and (**iii**) (20 nm) on (110)-oriented (R') $LaAlO_3$ substrates. All scans are 2 x 2 µm². (**e**) Calculated total internal energy of the T' and R' phases as a function of miscut angle. (**f**) DFT calculated (lines) and experimental (data points) pseudocubic

lattice parameters as a function of miscut angle. In (f), error bars for the symbols (from calculation of lattice parameters) are smaller than the symbol size. In (**b**) and (**e**), the energy shown is with respect to that of the *R*3*c* ground state.

The logic of our approach is given in **Fig. 1b**. Here, the free energy of the T' and R' polymorphs as a function of applied strain for different miscut angles was calculated using density functional theory (DFT; see ***Methods***). **Fig. 1c,d** offer comparisons of the crystallographic (**Fig. 1c**) and resulting topographical (**Fig. 1d**) features for each ground state BFO polymorph, when grown on LAO. First, for an exact (100) substrate (zero miscut), the local minimum at -5% [**Fig. 1b(i)**] implies that the T' phase of BFO [**Fig. 1d(i)**] is the ground state under compressive strains larger than ~4.5%. Increasing the film thickness to ~50 nm results in a phase coexistence of R'/T' needles embedded within a T' matrix [**Fig. 1d(i)**]. Next, in the (110) orientation (45-degree miscut) [**Fig. 1b(iii)**] the T' curve's minimum is centered to the right of its intersection with the R' curve. This suggests that the T' phase is energetically unfavourable for any practically achievable strains, consistent with experimental observations that the R' phase [**Fig. 1c(iii)**] is typically observed throughout the film [**Fig. 1d(iii)**]. Finally, for the intermediate (310) substrate orientation (~18.4-degree miscut from the (100) plane towards the [010] direction, **Fig. 1c(ii)**) is characterized by a considerable flattening of the energy landscape [**Fig. 1b(ii)**]. This flattening is evidenced by the shallow intersection (almost a straight line) between the free energy curves, as compared with the considerably sharper intersection angle of the T' and R' curves for zero miscut [**Fig. 1b(i)**]. Another key difference between conventional orientations is the almost filament-like contours exhibited in the (310)-oriented BFO topography, with no evidence of phase separation [**Fig. 1d(ii)**].

Our DFT calculations also show that at strain levels imposed by LAO, the transition between the R' and T' polymorphs occurs at a miscut angle close to 20 degrees (**Fig. 1e**). Therefore, a BFO film grown on LAO with a miscut near this value would be "on the brink," allowing the system to straddle the boundary between the polymorphs. Our approach to use (310) LAO substrates uniquely provides the perfect combination of perturbations: first, compressive epitaxial strain, which increases the axial ratio (c/a) and tilts the polar axis towards [001] (Ref. [36]); second, in-plane anisotropy, which allows access to average strain levels impossible with no miscut; and third, a crystallographic tilt which stabilizes low crystal symmetries that allow polarization rotation.

**Figure 1f** demonstrates that such hypothesis is supported by both experimental and computational results. We grew 20 nm thick epitaxial films of BFO on $(100)_{pc}$-, $(110)_{pc}$-, and $(310)_{pc}$-oriented LAO (nominal misfit of -4.5%) substrates by pulsed laser deposition (***Methods***) (pc = pseudocubic). The pc lattice parameters of the three films measured using x-ray diffraction (XRD) (***Methods***) are presented in **Fig. 1f** (data points). At a miscut of 0 degrees, the *a* and *b* lattice parameters are fixed to that of the LAO, while the *c* parameter is ~4.65 Å, characteristic of the T' phase[19]. For a miscut of 45 degrees, the *a*, *b*, and *c* lattice parameters are almost equal, indicative of the R' phase. Our DFT calculations (lines in **Fig. 1f**) (*Methods*) reproduce this trend, and also reveal that going from a miscut angle of 15 to 20 degrees, the *c* lattice parameter dramatically decreases, indicating the transition from the T' to R' phase. Indeed this is observed experimentally: at miscut of ~18.4 degrees [(310) orientation], the BFO film's lattice parameters are halfway between the R' and T' polymorphs: the *c* parameter is ~4.30 Å, the *a* lattice parameter for BFO remains clamped to the LAO substrate, while along the orthogonal $[1\bar{3}0]_{pc}$ in-plane direction the film is partially relaxed. This anisotropic strain condition causes the symmetry to become triclinic (*Supplementary Note 1*). We call this phase *super-R'* (s-R') BFO since its average strain in the *ab* plane ($\approx$ 3.7%) is the highest of any pure R' polymorph of BFO.

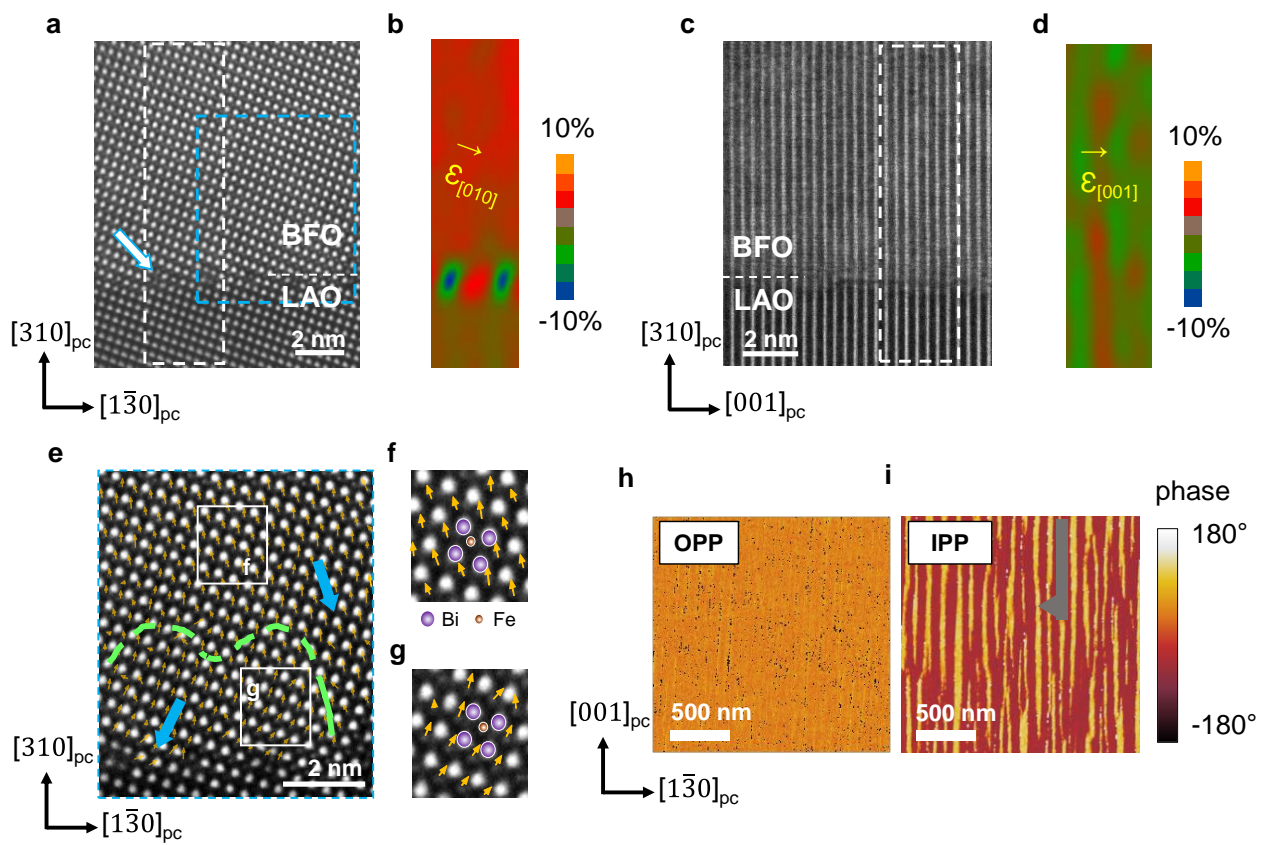

**Figure 2. Scanning transmission electron microscopy (STEM) and piezoresponse microscopy characterization of a super-R BiFeO₃ film.** (**a**) Atomic resolution STEM image along the $[001]_{pc}$ zone axis (the arrow points the location of a misfit dislocation at the interface), and (**b**) strain map of the dashed white box in (a). (**c**) Atomic resolution STEM image along the $[1\bar{3}0]_{pc}$ zone axis, and (**d**) strain map of dashed white box in (c). (**e**) Magnified view of (a) inside the blue dashed box. The blue arrows indicate the net polarization rotation across the green dashed line. (**f**) and (**g**) are magnified views of (e) inside the corresponding white boxes, clearly showing the atomic displacements in each region. The yellow arrows show the direction of the displacement of the Fe ions within the Bi cages. Piezoresponse force microscopy (taken on standard Asylum Cypher): (**h**) out-of-plane phase (OPP) and (**i**) in-plane phase (IPP) images. Both images are 2 x 2 µm².

The crystallography of the s-R' BFO on LAO (310) was further examined using high-angle annular dark-field scanning transmission electron microscopy (HAADF-STEM). A STEM image taken along the [001]$_{pc}$ zone axis (**Fig. 2a**) shows the clear miscut of ~18.4 degrees and several misfit dislocations at the substrate-film interface. The dislocation cores are clearly observed in the geometrical phase analysis (GPA) strain map[37] of the boxed region (**Fig. 2b**) (full strain maps and selected area electron diffraction are given in ***Supplementary Note 2***). The STEM image collected along the orthogonal zone axis [1$\bar{3}$0]$_{pc}$ (**Fig. 2c**) does not resolve individual atoms due to the close-packed positions of the atoms. However, the corresponding GPA strain map (**Fig. 2d**) shows that the film is coherently strained along the [001]$_{pc}$ in-plane direction. The ferroelectric structure was also attainable from the STEM images: a map of the local polarization (***P***) direction (**Fig. 2e**), extracted from the Fe atom displacement within the Bi cages from **Fig. 2a**, reveals the presence of two domains with different polarization directions, *i.e.* left-down and right-down (**Fig. 2f,g**). Piezoresponse force microscopy (PFM) confirms this: the as-grown ferroelectric domain arrangement comprises in-plane stripes, with uniform out-of-plane polarization (**Fig. 2h**), and two domain variants along the [1$\bar{3}$0]$_{pc}$ in-plane direction (**Fig. 2i**).

The structural analyses from XRD and STEM results thus indicate that the s-R' film is fully strained to the substrate along [001]$_{pc}$, while along [1$\bar{3}$0]$_{pc}$ it is partially relaxed. It is precisely this delicate balance of unequal strains imposed along each in-plane direction that distinguishes our approach from conventional strain engineering and prevents phase separation into T' and R' polymorphs. This allows us to distort the R' polymorph to its physical limits, *i.e.* to adopt the s-R' phase, which is not possible using biaxial strain.

Having established the s-R' phase crystallography, we now discuss its electromechanical response, which we measured on a sample with a thin (~2 nm) layer of La$_{0.67}$Sr$_{0.33}$MnO$_3$ (LSMO) between the LAO substrate and the BFO film, which acts as a bottom electrode. The presence of the lower electrode layer induces relaxation in the BFO layer, such that a ~ 40% / 60% phase mixture of s-R' / R' is formed. The XRD structural characterization of this mixed phase sample is presented in ***Supplementary Note 1***. Before proceeding, we wish to point out that obtaining a true and reliable measurement of the electromechanical response (and associated d$_{33}$) for very thin films is exceedingly difficult. One of the best currently available methods is through an atomic force microscope (AFM) fitted with an interferometric displacement sensor (IDS) (***Methods***). This scanning probe technique uses laser interferometry to measure the precise displacement of the tip, and critically, unlike traditional optical beam-based PFM measurements of d$_{33}$ values, when correctly implemented, is mostly unaffected by electrostatic contributions[38]. In our crystal geometry, the electric field was applied along [013]$_f$ ([310]$_{pc}$) (**Fig. 3f**) (subscript 'f' denotes 'film'; see ***Supplementary Note 1***), and the piezoresponse was measured along this direction. Since this is not strictly along the '3' direction of the film's crystal structure, we use d'$_{33}$ to denote this out-of-plane piezoresponse. The IDS data were collected on a 30 x 30 grid of points over a 2 x 2 µm$^2$ region of the film. Here we focus on a representative spatial location with high response; the full data for all spatial locations are presented in **Supplementary Note 4**. **Fig. 3a** shows the piezoelectric displacement (amplitude in pm/V as the amplitude of the applied ac voltage is 1 V) as a function of applied bias, with the corresponding phase data. In both datasets, the ON and OFF field labels correspond to a measurement *with* and *without* DC bias applied, respectively. The phase data for ON and OFF field are nominally identical, showing a 180° change indicative of ferroelectric switching, but the amplitude signal shows a clear difference between the OFF-field and ON-field measurements. While the amplitude of this piezoresponse for the OFF-field case (~60 pm/V) is consistent with previously-reported values for R-like BFO (**Supplementary Note 8**), the response shown for the ON-field case (approaching ~200 pm/V) is among the largest values reported for an epitaxially clamped $BiFeO_3$-based thin film, and is comparable to clamped industry standard PZT films[39]. We point out here that this value is only valid in this geometry and is likely to be different from the equivalent bulk value for this material (if such a measurement were possible). Also, such a value should be interpreted with caution, given the sample roughness; however, we can say with confidence that a significant enhancement of the piezoresponse is present during the field on measurements (further details on the experimental approach in **Supplementary Note 4**). Moreover, even in the OFF field state, a measurement of 60 pm/V for a 20 nm thick film is itself rather remarkable.

We have mapped out the piezoresponse over a 2 x 2 $\mu m^2$ region of the film, with the results summarized in Fig. 3b,c. The contour map (Fig. 3b) shows that in some regions, the measured ratio of the field ON to field OFF $d'_{33}$ maximum values reaches 4; we hypothesize that this corresponds to spatial locations where the super R phase is prevalent. Further statistical analysis (Fig. 3c) shows that for the 900 loops measured, the mean of the maximum $d'_{33}$ recorded for field ON is 65 pm/V, while for field OFF the mean is 33 pm/V. This enhancement is likely the result of an electric field-induced phase transition, which we discuss in detail next. Considering the film thickness of 20 nm, a surface displacement of 200 pm translates to a net reversible strain of close to 1%, a large figure of merit for an epitaxially clamped film.

To show that this dramatic enhancement in piezoresponse is the result of a field-induced phase transition, we used band-excitation piezoresponse spectroscopy (BEPS)[40] (**Methods**), once again on a 2 x 2 $\mu m^2$ region of 30 x 30 points. BEPS measures the cantilever response around a band of frequencies with the tip in contact with the sample while a bias waveform is applied. At each DC bias voltage step in the triangle wave (**Fig. 3d**), the frequency dependence of the electromechanical response of the sample is measured, yielding information about changes in the resonance. A systematic and reproducible change in this contact resonance frequency is strongly suggestive of a modification of the elastic properties of the material[41]. This technique has been successfully employed to demonstrate elastic softening at R/T phase boundaries in BFO thin films caused by an electric field-induced transition[41].

A representative BEPS spectrogram of the 20 nm s-R' / R' film taken at a single spatial location, magnified around the field-induced phase transition (**Fig. 3e**) reveals that during the ON-field bias application there is a distinct decrease (*i.e., softening*) of the resonant frequency (of around 5 kHz or 0.7%), suggesting a change in the elastic modulus. In the context of previous reports[41,42], we interpret this shift as an indication of a field-induced phase transition (see schematic of proposed phase transition in **Fig. 3f**). The spatial mapping of BEPS (measured over 2 x 2 µm²) shows that the elastic softening behaviour is observed everywhere (***Supplementary Note 5***), with some spatial variation in the degree of softening, as observed in the IDS measurements above. Moreover, the fatigue testing of the sample shows that ferroelectric switching occurs for > 10,000 driving cycles (***Supplementary Note 5***); however, the softening behaviour does decay after several hundred cycles.

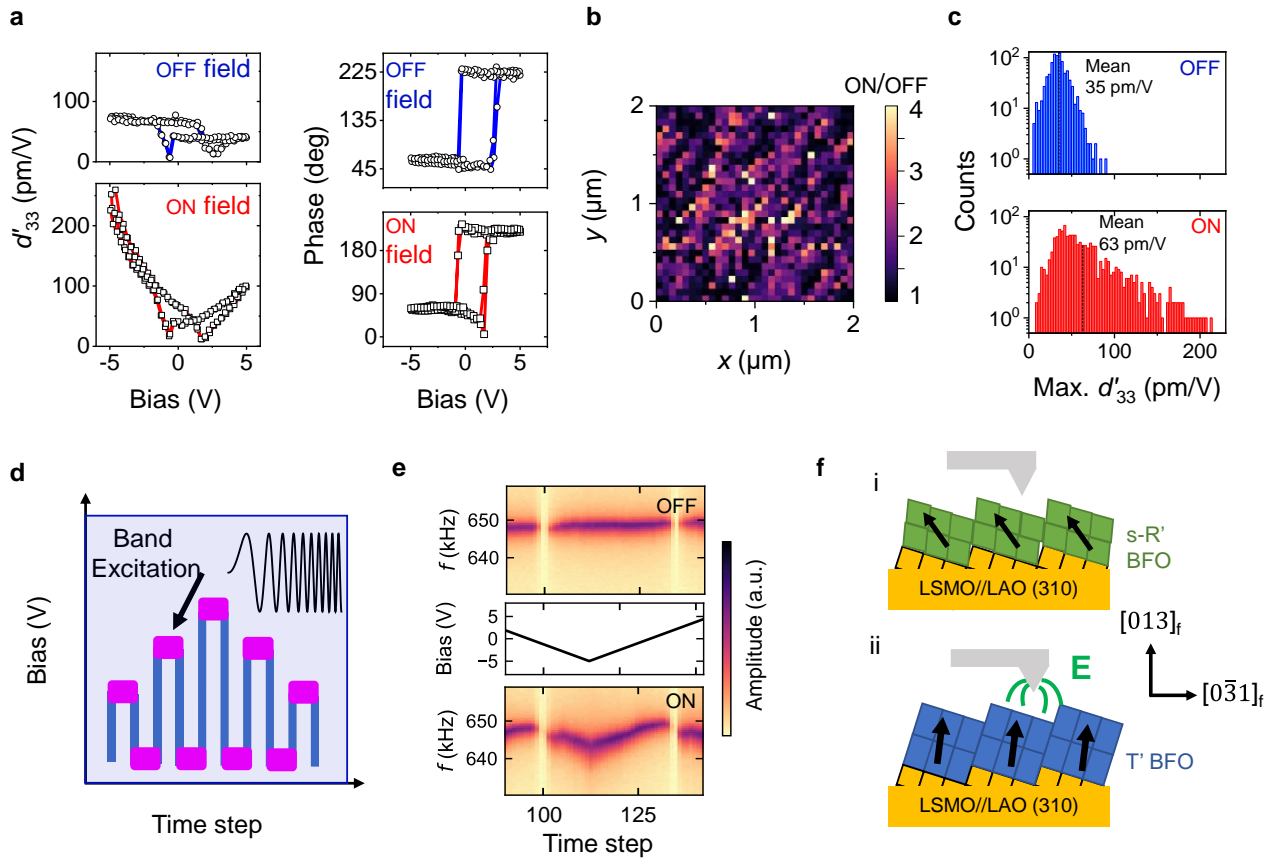

**Figure 3. Interferometric displacement sensor (IDS) and band excitation piezoresponse spectroscopy (BEPS) measurements.** (**a**) Selected representative piezoresponse amplitude and phase for IDS field OFF and field ON measurements for a 20 nm thick mixed super-R / R' BFO film. (**b**) spatial map of the ratio of measured d'$_{33}$ for field on to d'$_{33}$ for field off, showing some regions where the enhancement is more than 4 times. (**c**) histograms of the maximum d'$_{33}$ value for field on (top) and field off (bottom). (**d**) Sketch of the triangle wave for the BEPS measurement in (e), where the pink regions indicate the band-excitation frequency sweep. (**e**) BEPS for (top) field OFF and (bottom) field ON

states, showing evidence of elastic softening. (**f**) schematic of field-induced phase transition: (**i**) for low field (OFF), the film comprises the s-R' phase, (**ii**), at field ON, the film converts to the T' like phase.

To gain further insight into the physical properties of s-R' BFO, particularly regarding the field-induced phase transition, 0 K first-principles DFT (ref. [43]) and 300 K first-principles-based effective Hamiltonian[44] ($H_{eff}$) calculations were performed (***Methods***). Here, for clarity, we focus on the DFT results. $H_{eff}$ calculations yielded qualitatively similar results; see ***Supplementary Note 7*** for details.

Panels **a,b,c** in **Fig. 4** present the DFT computed properties related to polarization and octahedral tilting as a function of applied electric field, while panels **d,e,f** focus on the electromechanical and elastic response for comparison with measurements shown in **Fig. 3**. We first consider the zero-field state: here DFT finds a triclinic (*P*1) structure, characterized by non-equal components of the polarization vector ***P*** along the pseudocubic axes ($p_x$, $p_y$, $p_z$) (**Fig. 4a**). These computed polarization components allow us to deduce the angle subtended by ***P*** with the [001]$_f$ direction (**Fig. 4b**). Another result of the strongly distorted triclinic structure is the finding that the octahedral rotations ($\omega_x$, $\omega_y$, $\omega_z$) (**Fig. 4c**) are unequal.

With regard to the electromechanical properties at zero field, our computations show that the s-R' phase possesses a $d'_{33}$ of ~110 pm/V (full literature survey of piezoelectric coefficient values for BFO in ***Supplementary Note 8***). The computed elastic stiffness $C_{33}$ of the s-R' phase at zero field (**Fig. 4f**) is considerably lower than that of the *R*3*c* phase (129-150 GPa; Ref. [45]), indicating that it is intrinsically mechanically soft. The $C_{33}$ is also comparable to other computations for T' and R' BFO (refs. [45,46]).

Our calculations under applied field (**Fig. 4**) show two regions distinguished by a critical field $E_{cr}$ ~1.7 MV/cm, at which large changes in physical properties are observed. Upon increasing field, the polarization components (octahedral tilting) shown in **Fig. 4a** (**Fig. 4c**) systematically increase (decrease) until the abrupt transition at $E_{cr}$. Here, the in-plane components of the polarization decrease, the octahedral rotations are completely suppressed along the *z* direction and are reduced along the other two directions. This transition thus suggests a conversion from the s-R' BFO phase to the T' phase (still with a slight triclinic distortion). Notably, at $E_{cr}$ the polarization undergoes a dramatic rotation of ~8 degrees (**Fig. 4b**). This field-induced transition is schematically illustrated in **Fig. 3e**.

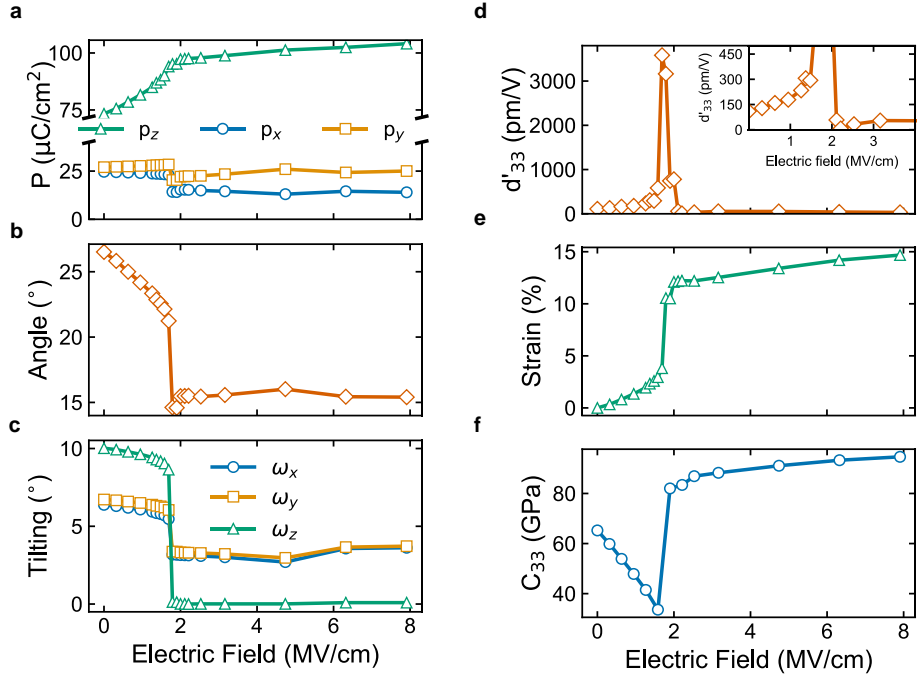

**Figure 4. Density functional theory predictions of properties as a function of applied electric field along [310]$_{pc}$ (≈ [013]$_f$).** (**a**) Components of the polarization ***P***, (**b**) angle subtended by ***P*** with [001]$_f$, and (**c**) octahedral tilting angles for a super-R film under applied field. (**d**) Piezoelectric coefficient d'$_{33}$, (**e**) total strain along [310]$_{pc}$ (≈ [013]$_f$), and (**f**) elastic constant C'$_{33}$ of the super-R BFO phase, as a function of applied electric field.

These structural changes are accompanied by anomalous signatures in the electromechanical coefficients. The d'$_{33}$ diverges approaching the transition (with values up to 3000 pm/V) and is accompanied by an abrupt increase in strain (**Fig. 4e**). Additionally, the elastic coefficient C'$_{33}$ (**Fig. 4f**) systematically decreases upon approach to E$_{cr}$, above which it rebounds and then continues to increase towards saturation.

Combining theoretical and experimental results, we thus draw key inferences vis-à-vis the implications of the elastic softening observed in the s-R' BFO system. Elastic anomalies located at a phase transition have been observed for various materials[47,48,49], the origins of which may be the softening of acoustic phonon modes triggered by external electric field[41]. One possible mechanism is the instability of the lower symmetry phase (in our case s-R') under increasing electric field. At some critical field, it becomes energetically favourable to soften the relevant elastic moduli along the field direction, thus accommodating the associated strain related to the interconversion between s-R' and T' phases[47,49].

An alternate viewpoint associates the phase coexistence to an anomalous softening[50]. In this picture, within the vicinity of a phase boundary, a vanishing polarization anisotropy dramatically decreases the domain wall energy. This phenomenon (observed in Sm-doped BFO thin films[51]), comes from the idea that a flattening of the free energy curve allows the polarization to easily rotate (*cf.* introduction). In the context of rotator and extender piezoelectrics given by Davis *et al.*[27], this would correspond to the former. Since we apply the electric field along a "arbitrary" non-polar direction, we invoke shear displacement as opposed to a simple extension along the c-axis. Such a spatial degeneracy of the polar axis implies that the polarization anisotropy energy must significantly weaken, decoupling the polar axis from the crystal lattice[50], and leading to anomalous softening.

The ability to switch from s-R' to T' with electric field demonstrates a handle on controlling the elastic properties as a result of the phase boundary that is accessed through a careful balance of strain and anisotropy. This undoubtedly provides various device opportunities, for instance, related to the giant increase in magnetoelectric properties predicted to coincide with structural softness in BFO (ref. [52]). Another exciting aspect relates to the formation of emergent ferroelectric topologies: the capacity to controllably decouple the polarization from the crystal lattice thus makes possible engineered rotational (*i.e.* twists) and chiral domain walls which may not be permitted by the bulk symmetry[53].

Our approach of anisotropic epitaxy using such highly miscut substrates has much broader perspectives than simply BFO and ferroelectric materials. For example, our pilot studies on magnetic oxides suggest that carefully optimised orientations of $La_{0.67}Sr_{0.33}MnO_3$ can reveal giant modulations in magnetic anisotropy[54]. One can also envisage with optically active oxides[55] (including BFO and other ferroelectrics such as $BaTiO_3$, Ref. [56]) that such low symmetry phases will harbour dramatically enhanced or modified optical responses, particularly with regard to anisotropy and *e.g.* photocurrents[57]. Finally, but not negligibly, the strongly anisotropic nature of the strain applied in such miscut substrates lends itself to novel strain relaxation mechanisms and thus one can expect that emergent flexoelectric (polarization induced by strain gradient) phenomena[58] await to be discovered.

In summary, we have used anisotropic epitaxy (AE) with highly miscut substrates to craft a mechanically soft super-R' triclinic phase of BFO. This polymorph is not energetically favourable under isotropic strain and hence cannot be obtained under nominal epitaxial conditions. This new phase, through its intrinsically elastically susceptible nature, shows a dramatic increase in the electromechanical response under an applied electric field. While as-grown super-R' exhibits a piezoelectric $d'_{33}$ value of 70 pm/V, an applied electric field induces a s-R' to T' phase transition which raises the piezoresponse by more than 200% and results in $d'_{33}$ values comparable to fully-clamped epitaxial PZT films. We also show that this field-induced phase transition yields reversible strain of up to 1%. Moreover, these effects are possible at modest driving voltages, making this very attractive for future low-energy electromechanical devices. Given recent breakthrough advances in 2D oxide synthesis techniques, free-standing films fabricated from templates using the AE approach could potentially offer super-elastic properties and record-breaking electromechanical performance. Further, as a generic heteroepitaxy technique it can offer myriad benefits for ferroelectric material systems. Wider perspectives of the AE technique encompass concepts such as giant anisotropy in low-symmetry magnetic oxides, and epitaxially metastable phases of multiferroics which could be harnessed in next-generation ferroelectric, piezoelectric, multiferroic, and spintronic device architectures.

**Methods**

**Epitaxial thin film fabrication:** BiFeO$_3$ thin films were grown by pulsed laser deposition using a KrF excimer laser with wavelength 248 nm. The LaAlO$_3$ substrates (110, 210, 310, 001) with 0.5 mm thickness were purchased from Shinkosha (Japan) and were single side polished with no specific surface treatment. The substrates were heated to 590°C and the films grown in an oxygen partial pressure of 100 mTorr. The laser fluence was 1-2 J/cm$^2$, and the films were cooled to room temperature at 20°C/min in 5 Torr of oxygen. To enable IDS measurements, a 2 nm-thick bottom electrode of conductive La$_{0.67}$Sr$_{0.33}$MnO$_3$ was inserted between the LAO (310) substrate and the BFO film. The presence of this electrode layer modifies the stability regions of the s-R' phase such that a phase mixture of s-R' and R' is unavoidable; however, through careful tuning of growth parameters and electrode thickness, we obtained a ~40%:60% volume fraction between s-R'- and R' phases for a nominal film thickness of 20 nm. XRD characterization of the mixed phase sample is given in *Supplementary Note 1*.

**Structural characterization:** X-ray diffraction was performed using Cu K$_{\alpha-1}$ radiation in a 9-kW rotating anode Rigaku SmartLab diffractometer and a PANAlytical Materials Research Diffractometer (MRD) systems. The lattice parameters of the pure s-R' BFO and the mixed s-R' / R' films were determined by measuring reciprocal space maps (RSMs) around various reflections, as detailed in *Supplementary Note 1*.

**Scanning transmission electron microscopy (STEM):** The STEM specimens were prepared by tripod polishing followed by final cleaning with a Getan precision ion polishing system (PIPS). The STEM images were acquired by double-corrected FEI Titan$^3$ 80-300 FEGTEM at 300 kV voltage with the convergence angle of 21 mrad.

**Ferroelectric domain characterization and interferometric displacement sensor measurements:** Standard PFM measurements were performed on a commercial atomic force microscope (Cypher, Asylum Research), to determine the as-grown in-plane and out of plane domain arrangements for both the pure s-R' film (grown directly on LAO 310) and the mixed s-R' / R' film (grown on LSMO-buffered LAO 310). The tips were Pt/Cr coated probes (ElectiMulti 75G, BudgetSensors) and the excitation voltage 500 mV for the in-plane and out-of-plane responses. Further PFM was carried out in ambient environment on a commercial atomic force microscope (Cypher AFM, Asylum Research), equipped with an in-house developed band-excitation controller based on a National Instruments PXIe-6124 data acquisition card operating in Labview software. Pt/Ir coated conductive Si probes (BudgetSensors ElectriMulti75-G) were used for all measurements. The interferometric measurements were performed on the same Cypher AFM with an integrated quantitative laser doppler vibrometer (LDV) system (Polytec GmbH, Waldbronn, Germany).

**Theoretical calculations:** DFT calculations were performed using the Vienna *ab-initio* simulation package (VASP)[43]. The revised Perdew, Burke, and Ernzerhof functional for solids (PBEsol)[59] was adopted, with Bi 6s6p, Fe 3d4s and O 2s2p electrons considered as valence electrons. A typical effective Hubbard $U$ parameter of 4 eV for the localized 3$d$ electrons of Fe ions[60]. The $k$-point mesh of 3×3×3 was employed for the 2×2×2 supercell. The plane wave energy cutoff was chosen to be 500 eV, and G-type antiferromagnetism was imposed. Constrained structural optimization was performed, as the in-plane lattice vectors were frozen to match the experimental strain values, while the out-of-plane vector and all atomic positions were free to relax, until all the Hellmann-Feynman forces converged to be smaller than 0.001 eV/Å on each ion. Starting with the *R3c* phase but removing symmetry, the R', s-R' and T' phases were obtained from structural optimization at the different investigated strains. The optimized structures of s-R' and T' phases, which corresponds to those in Fig. 4, are tabulated in ***Supplementary Note 6***. For **Fig. 1b**, various constraints on the lattice parameters were imposed, depending on the miscut. For 0-degree miscut, the in-plane $a$ and $b$ BFO lattice parameters were swept together, consistent with biaxial in plane strain. This yielded energy curves for T' and R' phases similar to those reported in various works[16,52]. For a miscut of 18.4 degrees, consistent with the experiment, the strain along the $a$ lattice direction was fixed at -4.7% and the $b$ lattice parameter was swept to map out the energy curves. Finally, for 45-degree miscut, we fixed the strain along $a$ to be consistent with experiment and swept the $b$ lattice parameter. The strain values given in **Fig. 1b** are the strain along $b$. These different strain constraints are the reason behind the different minima for the R' curves in **Fig. 1b**. To supplement the DFT results, first-principles based effective Hamiltonian method[61] calculations of the super-R' BFO phase were performed at 300 K. These yielded qualitatively similar results to those obtained by DFT, and the details are presented in ***Supplementary Note 7***. In addition, the piezoelectric responses shown in Fig. 4 were calculated by applying electric field within the DFT calculations with the method proposed and adopted in Refs. [62,63]. Specifically, for d'$_{33}$, we apply an electric field $E$ along the [013] direction and optimize the structure; then the piezoelectric constant was obtained via d'$_{33} = \eta/E|_{[013]}$, where $\eta$ is the change of strain along the [013] direction. The phonon spectra shown in ***Supplementary Note 6*** are calculated with PHONOPY[64] using 4×4×4 supercells (that each contain 320 atoms) and a single k-point located at the zone center. Finally, the polarization was calculated using the modern (Berry-phase) theory of polarization[65].

**Acknowledgements**

This research was partially supported by the Australian Research Council Centre of Excellence in Future Low-Energy Electronics Technologies (Project No. CE170100039) and funded by the Australian Government. D.S. and V.N. acknowledge the support of the ARC through Discovery grants. O.P. acknowledges the Australian Government Research Training Program Scholarship, the Australian Institute for Nuclear Science and Engineering (AINSE) Post-graduate research award, and the Scholarship AINSE ANSTO French Embassy (SAAFE) program. C.X. and L.B. thank the DARPA Grant No. HR0011727183-D18AP00010 (TEE Program). X. C acknowledges the use of facilities within the Monash Centre for Electron Microscopy (MCEM), which were funded by Australian Research Council grant ARC Funding (LE0454166) and ARC Funding (LE0882821). The piezoresponse spectroscopy measurements were supported by the U.S. Department of Energy (DOE), Office of Science, Basic Energy Sciences (BES), Materials Sciences and Engineering Division (K.K., R.K.V., L.C.) and was performed at the Oak Ridge National Laboratory's Center for Nanophase Materials Sciences (CNMS), a U.S. DOE Office of Science User Facility. We thank C. Paillard for fruitful discussions and for sharing his scripts.

**Author contributions**

D.S. and N.V. conceived and supervised the study. O.P. and D.S. fabricated the films and performed XRD experiments and analysis. X.C., Y.Z., and A.d.M carried out STEM and GPA analysis. R.K.V., K.K., and L.C. performed the band-excitation piezoresponse spectroscopy and interferometric displacement sensor measurements and analysed the data. C.X. performed DFT calculations and B.X. carried out effective Hamiltonian simulations under the supervision of L.B. O.P., D.S., and V.N. wrote the manuscript. All authors contributed to data analysis, manuscript preparation, and commented on the manuscript.

**Competing Interests**

The authors declare no competing financial interests.

**Data Availability**

The data that support the findings of this study are available from the corresponding author upon reasonable request.

**References**

1. Fu, H. & Cohen, R. E. Polarization rotation mechanism for ultrahigh electromechanical response in single-crystal piezoelectrics. *Nature* **403**, 281–283 (2000).

2. Zhang, S. *et al.* Advantages and challenges of relaxor-PbTiO$_3$ ferroelectric crystals for electroacoustic transducers – A review. *Progress in Materials Science* **68**, 1–66 (2015).

3. Damjanovic, D. Contributions to the Piezoelectric Effect in Ferroelectric Single Crystals and Ceramics. *Journal of the American Ceramic Society* **88**, 2663–2676 (2005).

4. Ahart, M. *et al.* Origin of morphotropic phase boundaries in ferroelectrics. *Nature* **451**, 545–548 (2008).

5. Cross, E. Lead-free at last. *Nature* **432**, 24–25 (2004).

6. Jaffe, B. *Piezoelectric Ceramics*. (Academic Press, 1971).

7. Li, F. *et al.* The origin of ultrahigh piezoelectricity in relaxor-ferroelectric solid solution crystals. *Nat Commun* **7**, 13807 (2016).

8. Liu, W. & Ren, X. Large Piezoelectric Effect in Pb-Free Ceramics. *Phys. Rev. Lett.* **103**, 257602 (2009).

9. Noheda, B. *et al.* Polarization Rotation via a Monoclinic Phase in the Piezoelectric 92%PbZn$_{1/3}$Nb$_{2/3}$O$_3$-8%PbTiO$_3$. *Phys. Rev. Lett.* **86**, 3891–3894 (2001).

10. Saito, Y. *et al.* Lead-free piezoceramics. *Nature* **432**, 84–87 (2004).

11. Schönau, K. A. *et al.* Nanodomain structure of Pb[Zr$_{1-x}$Ti$_x$]O$_3$ at its morphotropic phase boundary: Investigations from local to average structure. *Phys. Rev. B* **75**, 184117 (2007).

12. Zeches, R. J. *et al.* A Strain-Driven Morphotropic Phase Boundary in BiFeO$_3$. *Science* **326**, 977–980 (2009).

13. Jin, Y. M., Wang, Y. U., Khachaturyan, A. G., Li, J. F. & Viehland, D. Conformal miniaturization of domains with low domain-wall energy: Monoclinic ferroelectric states near the morphotropic phase boundaries. *Physical Review Letters* **91**, (2003).

14. Noheda, B. *et al.* A monoclinic ferroelectric phase in the Pb(Zr$_{1-x}$Ti$_x$)O$_3$ solid solution. *Appl. Phys. Lett.* **74**, 2059–2061 (1999).

15. Noheda, B. *et al.* Stability of the monoclinic phase in the ferroelectric perovskite PbZr$_{1-x}$Ti$_x$O$_3$. *Phys. Rev. B* **63**, 014103 (2000).

16. Sando, D., Xu, B., Bellaiche, L. & Nagarajan, V. A multiferroic on the brink: Uncovering the nuances of strain-induced transitions in BiFeO$_3$. *Appl. Phys. Rev.* **3**, 011106 (2016).

17. Sando, D. *et al.* Large elasto-optic effect and reversible electrochromism in multiferroic BiFeO$_3$. *Nature Communications* **7**, 10718 (2016).

18. Sando, D. *et al.* Crafting the magnonic and spintronic response of BiFeO$_3$ films by epitaxial strain. *Nature Mater* **12**, 641–646 (2013).

19. Béa, H. *et al.* Evidence for Room-Temperature Multiferroicity in a Compound with a Giant Axial Ratio. *Phys. Rev. Lett.* **102**, 217603 (2009).

20. Zhang, J. X. *et al.* Large field-induced strains in a lead-free piezoelectric material. *Nature Nanotechnology* **6**, 98–102 (2011).

21. Zhang, J. *et al.* A nanoscale shape memory oxide. *Nat Commun* **4**, 2768 (2013).

22. Edwards, D. *et al.* Giant resistive switching in mixed phase BiFeO$_3$ via phase population control. *Nanoscale* **10**, 17629–17637 (2018).

23. Heo, Y., Jang, B.-K., Kim, S. J., Yang, C.-H. & Seidel, J. Nanoscale Mechanical Softening of Morphotropic BiFeO$_3$. *Advanced Materials* **26**, 7568–7572 (2014).

24. Chen, Z. *et al.* Coexistence of ferroelectric triclinic phases in highly strained BiFeO$_3$ films. *Phys. Rev. B* **84**, 094116 (2011).

25. Bellaiche, L., García, A. & Vanderbilt, D. Finite-Temperature Properties of PbZr$_{1-x}$Ti$_x$O$_3$ Alloys from First Principles. *Phys. Rev. Lett.* **84**, 5427–5430 (2000).

26. Lisenkov, S., Rahmedov, D. & Bellaiche, L. Electric-Field-Induced Paths in Multiferroic BiFeO$_3$ from Atomistic Simulations. *Phys. Rev. Lett.* **103**, 047204 (2009).

27. Davis, M., Budimir, M., Damjanovic, D. & Setter, N. Rotator and extender ferroelectrics: Importance of the shear coefficient to the piezoelectric properties of domain-engineered crystals and ceramics. *Journal of Applied Physics* **101**, 054112 (2007).

28. Damodaran, A. R. *et al.* Nanoscale Structure and Mechanism for Enhanced Electromechanical Response of Highly Strained BiFeO$_3$ Thin Films. *Advanced Materials* **23**, 3170–3175 (2011).

29. Li, F. *et al.* Giant piezoelectricity of Sm-doped Pb(Mg$_{1/3}$Nb$_{2/3}$)O$_3$-PbTiO$_3$ single crystals. *Science* **364**, 264–268 (2019).

30. Ren, X. Large electric-field-induced strain in ferroelectric crystals by point-defect-mediated reversible domain switching. *Nature Mater* **3**, 91–94 (2004).

31. Sinsheimer, J. *et al.* Engineering Polarization Rotation in a Ferroelectric Superlattice. *Phys. Rev. Lett.* **109**, 167601 (2012).

32. Schlom, D. G. *et al.* Strain Tuning of Ferroelectric Thin Films. *Annu. Rev. Mater. Res.* **37**, 589–626 (2007).

33. Gui, Z. & Bellaiche, L. Tuning and optimizing properties of ferroelectric films grown on a single substrate: A first-principles-based study. *Phys. Rev. B* **91**, 020102 (2015).

34. Xu, R. *et al.* Ferroelectric polarization reversal via successive ferroelastic transitions. *Nature Mater* **14**, 79–86 (2015).

35. Yan, L., Cao, H., Li, J. & Viehland, D. Triclinic phase in tilted (001) oriented BiFeO$_3$ epitaxial thin films. *Appl. Phys. Lett.* **94**, 132901 (2009).

36. Jang, H. W. *et al.* Strain-Induced Polarization Rotation in Epitaxial (001) BiFeO$_3$ Thin Films. *Phys. Rev. Lett.* **101**, 107602 (2008).

37. Hÿtch, M. J., Snoeck, E. & Kilaas, R. Quantitative measurement of displacement and strain fields from HREM micrographs. *Ultramicroscopy* **74**, 131–146 (1998).

38. Labuda, A. & Proksch, R. Quantitative measurements of electromechanical response with a combined optical beam and interferometric atomic force microscope. *Applied Physics Letters* **106**, 253103 (2015).

39. Nagarajan, V. *et al.* Realizing intrinsic piezoresponse in epitaxial submicron lead zirconate titanate capacitors on Si. *Appl. Phys. Lett.* **81**, 4215–4217 (2002).

40. Jesse, S. & Kalinin, S. V. Band excitation in scanning probe microscopy: sines of change. *J. Phys. D: Appl. Phys.* **44**, 464006 (2011).

41. Li, Q. *et al.* Giant elastic tunability in strained BiFeO$_3$ near an electrically induced phase transition. *Nat Commun* **6**, 8985 (2015).

42. Li, Q. *et al.* Probing local bias-induced transitions using photothermal excitation contact resonance atomic force microscopy and voltage spectroscopy. *ACS Nano* **9**, 1848–1857 (2015).

43. Kresse, G. & Joubert, D. From ultrasoft pseudopotentials to the projector augmented-wave method. *Phys. Rev. B* **59**, 1758–1775 (1999).

44. Kornev, I. A., Lisenkov, S., Haumont, R., Dkhil, B. & Bellaiche, L. Finite-Temperature Properties of Multiferroic BiFeO$_3$. *Phys. Rev. Lett.* **99**, 227602 (2007).

45. Shang, S. L., Sheng, G., Wang, Y., Chen, L. Q. & Liu, Z. K. Elastic properties of cubic and rhombohedral BiFeO$_3$ from first-principles calculations. *Phys. Rev. B* **80**, 052102 (2009).

46. Dong, H., Chen, C., Wang, S., Duan, W. & Li, J. Elastic properties of tetragonal BiFeO$_3$ from first-principles calculations. *Appl. Phys. Lett.* **102**, 182905 (2013).

47. Singh, A. K. *et al.* Origin of high piezoelectric response of Pb(Zr$_x$Ti$_{1-x}$)O$_3$ at the morphotropic phase boundary: Role of elastic instability. *Appl. Phys. Lett.* **92**, 022910 (2008).

48. Carpenter, M. A. & Salje, E. K. H. Elastic anomalies in minerals due to structural phase transitions. *ejm* **10**, 693–812 (1998).

49. Mishra, S. K., Pandey, D. & Singh, A. P. Effect of phase coexistence at morphotropic phase boundary on the properties of Pb(Zr$_x$Ti$_{1-x}$)O$_3$ ceramics. *Appl. Phys. Lett.* **69**, 1707–1709 (1996).

50. Rossetti, G. A., Khachaturyan, A. G., Akcay, G. & Ni, Y. Ferroelectric solid solutions with morphotropic boundaries: Vanishing polarization anisotropy, adaptive, polar glass, and two-phase states. *Journal of Applied Physics* **103**, 114113 (2008).

51. Borisevich, A. Y. *et al.* Atomic-scale evolution of modulated phases at the ferroelectric–antiferroelectric morphotropic phase boundary controlled by flexoelectric interaction. *Nat Commun* **3**, 775 (2012).

52. Wojdeł, J. C. & Íñiguez, J. Ab Initio Indications for Giant Magnetoelectric Effects Driven by Structural Softness. *Phys. Rev. Lett.* **105**, 037208 (2010).

53. Zhang, Q. *et al.* Nanoscale Bubble Domains and Topological Transitions in Ultrathin Ferroelectric Films. *Advanced Materials* **29**, 1702375 (2017).

54. Paull, O. in preparation.

55. Sando, D. *et al.* Linear electro-optic effect in multiferroic BiFeO 3 thin films. *Phys. Rev. B* **89**, 195106 (2014).

56. Eltes, F. *et al.* An integrated optical modulator operating at cryogenic temperatures. *Nature Materials* **19**, 1164–1168 (2020).

57. Bhatnagar, A., Roy Chaudhuri, A., Heon Kim, Y., Hesse, D. & Alexe, M. Role of domain walls in the abnormal photovoltaic effect in BiFeO 3. *Nature Communications* **4**, 2835 (2013).

58. Catalan, G. *et al.* Flexoelectric rotation of polarization in ferroelectric thin films. *Nature Mater* **10**, 963–967 (2011).

59. Perdew, J. P. *et al.* Restoring the Density-Gradient Expansion for Exchange in Solids and Surfaces. *Phys. Rev. Lett.* **100**, 136406 (2008).

60. Xu, C., Xu, B., Dupé, B. & Bellaiche, L. Magnetic interactions in $BiFeO_3$: A first-principles study. *Phys. Rev. B* **99**, 104420 (2019).

61. Prosandeev, S., Wang, D., Ren, W., Íñiguez, J. & Bellaiche, L. Novel Nanoscale Twinned Phases in Perovskite Oxides. *Advanced Functional Materials* **23**, 234–240 (2013).

62. Fu, H. & Bellaiche, L. First-Principles Determination of Electromechanical Responses of Solids under Finite Electric Fields. *Phys. Rev. Lett.* **91**, 057601 (2003).

63. Xu, C. *et al.* Electric-Field Switching of Magnetic Topological Charge in Type-I Multiferroics. *Phys. Rev. Lett.* **125**, 037203 (2020).

64. Chaput, L., Togo, A., Tanaka, I. & Hug, G. Phonon-phonon interactions in transition metals. *Phys. Rev. B* **84**, 094302 (2011).

65. King-Smith, R. D. & Vanderbilt, D. Theory of polarization of crystalline solids. *Phys. Rev. B* **47**, 1651–1654 (1993).